\documentstyle[12pt,epsfig]{article}
\let\section=\subsection  \let\subsection=\subsubsection

\def\be{\begin{equation}}
\def\ee{\end{equation}}
\def\bea{\begin{eqnarray}}
\def\eea{\end{eqnarray}}
 
\def\br{\mbox{\boldmath $r$}}
\def\bm{\mbox{\boldmath $m$}}

\begin{document}
\begin{center}
{\large \bf Bag Formation in Quantum Hall Ferromagnets.
}\\[5mm]
H. Walliser \\[5mm]
{\small \it 
Fachbereich Physik, Universit\"at Siegen,  
D57068 Siegen, Germany} 
\end{center}

\begin{abstract}\noindent

Charged skyrmions or spin--textures in the quantum Hall ferromagnet
at filling factor $\nu=1$ are reinvestigated using the Hartree--Fock
method in the lowest Landau level approximation. It is shown that
the single Slater determinant with the minimum energy in the unit charge
sector is always of the hedgehog form. 
It is observed that the magnetization vector's length deviates
locally from unity, i.e. a bag is formed which accommodates the
excess charge. In terms of a gradient expansion for extended
spin--textures a novel $O(3)$ type of effective action is presented,
which takes bag formation into account.

\end{abstract} 

\bigskip
\leftline{PACS 73.40.Hm, 73.20.Dx, 12.39.Dc}  
\leftline{Keywords: skyrmions, $O(3)$-model}

%\newpage

\section{Introduction}

Quantum Hall ferromagnets at filling factor $\nu=1$ exhibit
charged excitations that are topologically stable objects
called skyrmions. Various experiments \cite{sepw95,ssplck00}
have revealed strong evidence for the existence of these
objects in two--dimensional electron systems.

For their theoretical description we employ the
microscopic Hartree--Fock (HF) theory with the hedgehog
ansatz for charged spin--textures, developed by
Fertig et. al. \cite{fbcd94} and subsequently commonly
used in the literature \cite{c97,fbcdks97,apfgd97}.
We show that the most general single Slater determinant
in the lowest Landau level (LLL) approximation collapses
into the hedgehog ansatz, hence supporting this approach.
Physical observables like magnetization and charge 
density are then calculated microscopically in the 
standard way (section 2).
However in contrast to previous approaches, we will analyse the
full magnetization vector, not only its third component,
the spin--density.
It will turn out that the magnetization is not a unit
vector, but its length is substantially decreased in regions
with non--zero excess charge (bag formation). 
A signature of this behaviour
was already observed in \cite{apfgd97}, namely that the
absolute value of the spin--density at the origin as obtained
from the HF calculation is reduced compared to that of
the non--linear $O(3)$ model which is always unity.
However the conjecture that with quantum corrections
included, the discrepancies between effective field theory
and microscopic HF results may be resolved, is incorrect.
Bag formation appears already in tree level and accordingly
the magnetization vector can no longer be identified
directly with the $O(3)$ unit vector field. An additional
scalar field is necessary to describe the magnetization
vector's length.

Subsequently, using the gradient expansion, a 
new $O(3)$ type of effective action
for extended spin--textures is derived which takes these effects
into account (section 3). It will be demonstrated, that this
action reproduces the results of the microscopic HF calculation
correctly for zero range forces.

\section{Microscopic Hartree-Fock calculations}

As usual lengths are measured
in units of the magnetic length $\ell = \sqrt{\hbar c/eB}$
and energies in units of the Coulomb energy $e^2/\epsilon \ell$.
Then for a given electron-electron interaction in the LLL
there exists only one dimensionless parameter
in the theory, namely the effective Zeeman coupling 
$g=g_s \mu_B B /(e^2/\epsilon \ell)$.

In the completely filled LLL (filling factor $\nu=1$) the
ground state is given by the unique
single Slater determinant (Vandermonde's determinant, 
see e.g. \cite{g99})
\be\label{vander}
| \Psi \rangle = {\cal A} \prod_{m=0}^{\infty} | m \downarrow \rangle \ , 
\qquad \langle \br|m \rangle = \frac{1}{2\pi}  \frac{1}{2^m m!} \, 
r^m \, {\rm e}^{im\varphi}  {\rm e}^{- r^2/4}
\ee
with the spins of all single particle states down. Here we are
concerned with the description of charged spin-texture excitations,
in particular with those in the charge sectors $Q=\pm 1$. Due to
the particle-hole symmetry \cite{fbcdks97}
the energies of the positively charged
(quasi-hole, antiskyrmion) and negatively charged
(quasi-particle and skyrmion) excitations, $E^\pm = E \mp \epsilon_0$,
differ only by the exchange energy per electron in the 
ground state, e.g.
$\epsilon_0=-\sqrt{\pi/8}$ in the pure Coulomb case. Thus it is
sufficient to consider the $Q=+1$ sector with
the following ans\"atze for the many-body wave function:\\
(i) the quasi-hole
\be\label{quasihole}
| \Psi \rangle = {\cal A} \prod_{m=0}^{\infty} 
| m+1 \downarrow \rangle \, , 
\enspace \quad \qquad \qquad \qquad \qquad \qquad \qquad \qquad
(\mbox{quasi-hole})
\ee
with just one electron removed from the completely filled LLL.
Its energy relative to the ground state is $E^+=g/2-2\epsilon_0$
($E=g/2-\epsilon_0$).\\
(ii) the hedgehog
\be\label{hedgehog}
| \Psi \rangle = {\cal A} \prod_{m=0}^{\infty} 
\left( u_m |m \uparrow \rangle +
v_m | m+1 \downarrow \rangle \right) \, , \quad u_m^2+v_m^2=1 \, , 
\qquad  (\mbox{hedgehog})
\ee
introduced by Fertig et. al. \cite{fbcd94}. 
For $u_m=0,\, v_m=1, \,\forall m$, 
this ansatz reduces to the quasi--hole (\ref{quasihole}).\\
(iii) the general single Slater determinant 
\be\label{SD}
| \Psi \rangle = {\cal A} \prod_\alpha | \alpha \rangle \, , \quad
| \alpha \rangle = \sum_{m \sigma} C^\alpha_{m\sigma} 
|m\sigma\rangle \, , 
\qquad \qquad \quad (\mbox{general single SD})
\ee
whose single particle states $|\alpha\rangle$ with $\langle
\alpha|\alpha\rangle =1$ are given by the most general
linear combination of LLL-states labeled by
magnetic quantum number $m=0,\ldots ,N-1$ and spin $\sigma$ 
(up and down).

A HF calculation for the general Slater determinant (ansatz (\ref{SD}))
was performed with $N-1$ particles
in order to guarantee $Q=+1$ in a basis up to
$N=300$ states involving the diagonalization of $600 \times 600$
matrices. Independently from the initial conditions chosen,
the general Slater determinant always collapsed
\be
C^\alpha_{m \downarrow} = u_\alpha \delta_{m,\alpha} \, , \qquad 
C^\alpha_{m \uparrow}   = v_\alpha \delta_{m,\alpha+1}
\ee 
into the specific hedgehog ansatz (\ref{hedgehog}) during the
iteration process. This happened for
various Zeeman coupling strengths $g$ and also for
different (repulsive) interactions. Thus, we may consider this
finding as a strong indication that the lowest energy single
Slater determinant for $Q=1$ in the LLL is indeed always of the
hedgehog form. I am not aware of a rigorous proof of that 
statement, although it is certainly related to the fact that
in the ansatz (\ref{hedgehog}) only single particle states 
with the same grand-spin $m+\sigma$ couple.

For the reasons just explained we may use the hedgehog ansatz
(\ref{hedgehog}) in the $Q=+1$ sector which simplifies the
calculation considerably. Instead of diagonalizing matrices
there are only coupled algebraic equations for
$u_m$ and $v_m$ to solve for each $m$ separately. We give
the expression for the HF energy derived in \cite{fbcd94}
relative to the completely filled LLL
\bea\label{hf}
E^+_{HF} &=& \frac{g}{2} \sum_m (u_m^2-v_{m-1}^2+1) \nonumber \\
&+& \frac{1}{2} \sum_{mm^\prime} (u_m^2+v_{m-1}^2-1)
  (u_{m^\prime}^2+v_{m^\prime-1}^2-1) \langle m\,m^\prime|V|
m\,m^\prime \rangle \nonumber \\
&-& \frac{1}{2} \sum_{mm^\prime} (u_m^2 u_{m^\prime}^2+
v_{m-1}^2 v_{m^\prime-1}^2 - 1) 
\langle m\,m^\prime|V|m^\prime\, m \rangle \\
&-& \sum_{mm^\prime} u_m v_m u_{m^\prime} v_{m^\prime} 
\langle m\,m^\prime+1|V|m^\prime\, m+1\rangle \, . \nonumber 
\eea
The individual terms are the Zeeman energy and the direct and
exchange terms of the
two-body interaction $V$. The corresponding HF equations are
immediately derived. The (excess) charge density 
\bea\label{charge}
\rho^C(\br) & \equiv & \frac{1}{2\pi} -
\langle \Psi|\sum_i \delta(\br-\br_i)|\Psi \rangle 
\nonumber \\
& = & \frac{1}{2\pi} \sum_{m} (1-u_m^2-v_{m-1}^2)
   \, \frac{x^m}{m!}\, {\rm e}^{-x} \, , \qquad x=\frac{r^2}{2}  
\eea
is radially symmetric and normalized to $Q^C=+1$ as may be 
checked using the condition $u_m^2+v_m^2=1$. The magnetization vector
\bea\label{magnet}
&&\bm (\br) \equiv -2\pi 
\langle \Psi|\sum_i \mbox{\boldmath $\sigma$}(i)
\delta(\br-\br_i)|\Psi \rangle 
=\left( f(r) \cos\varphi , f(r) \sin\varphi , h(r) \right)
\nonumber \\
&& f(r)=-\sqrt{2} \, r \sum_m \frac{u_m v_m}{\sqrt{m+1}} \, 
\frac{x^m}{m!} \, {\rm e}^{-x} \, , \qquad
h(r)=\sum_m (v_{m-1}^2-u_m^2) \, \frac{x^m}{m!}\, {\rm e}^{-x} \, ,
\eea
($\mbox{\boldmath $\sigma$}(i)$ are the Pauli matrices for 
particle $i$) normalized to $\lim_{r \to \infty} \bm =
\mbox{\boldmath $e$}_3$
is of the hedgehog form as expected. However, it is certainly not
a unit vector: there are two radial functions involved for which
in general $f^2+h^2\not=1$. In order to make closer contact to
effective field theories we may rewrite the magnetization vector
in the form
\be\label{magnet1}
\bm (\br) = \sigma(r) \hat{\bm} =
\sigma(r) \left( \sin F(r) \cos \varphi ,
\sin F(r) \sin \varphi , \cos F(r) \right) \, 
\ee
with $\hat{\bm}$ a unit vector. The scalar function $\sigma$
describes the deviation of the magnetization vector's length from
$1$ (bag formation) and the angle function $F$ determines the
orientation of the unit vector $\hat{\bm}$. It is clear that
these functions are readily expressed in terms of the original 
functions $f$ and $h$. 

In this way we are going to analyse the full magnetization vector
$\bm$, not only its third component, which is the main difference
to previous presentations. It will immediately lead us to
an important relation of the formed bag with the charge.
In order to better
understand the situation we will study the analytical hard-core
model before we present numerical results for the more
realistic Coulomb interaction.

\subsection{Hard-core interaction}

For a zero range interaction 
$V(\br_1 - \br_2) = t \delta (\br_1 - \br_2)$
and vanishing Zeeman splitting, $g=0$, the HF equations may be solved
analytically
\be\label{uv}
u_m=\sqrt{\frac{a}{m+1+a}} \, , \quad 
v_m=\sqrt{\frac{m+1}{m+1+a}} \, , \qquad a=\frac{\lambda^2}{2} \, . 
\ee
Here, $\lambda$ is a free parameter which fixes the soliton size.
The radial functions determining the magnetization $\bm$ in 
(\ref{magnet}) are
\be\label{fh}
f(r) =  -\lambda r \, p(x,a) \, , \quad
h(r) = (x-a) \, p(x,a) \, , \qquad x=\frac{r^2}{2}
\ee
with the single valued analytical function
\bea\label{gamma}
&& p(x,a) = \sum_m \frac{1}{m+1+a} \, \frac{x^m}{m!} \, 
{\rm e}^{-x} \, , \nonumber \\
&& x^{a+1} \, p(x,a) \, {\rm e}^x = \int_0^x dt \, 
t^a \, {\rm e}^t \, , \\
&& \frac{d}{dx} (x p(x,a)) + (a+x) p(x,a) = 1 \, ,
\nonumber
\eea
related to the incomplete gamma function \cite{as65}.
From (\ref{magnet1}) and (\ref{fh}) we obtain
\be\label{fs}
F(r) = 2 \arctan \frac{\lambda}{r} \, , \qquad
\sigma (r) = (x+a) \, p(x,a) 
\ee
the angle function of the familiar Belavin--Polyakov soliton
\cite{bp75}, but the 
modulus function $\sigma$ is not equal to $1$.
The topological and physical charge densities are
\bea
&& \rho^T(r) = \frac{1}{4\pi r} \frac{d}{dr} (\cos F)
= \frac{1}{\pi} \frac{\lambda^2}{(r^2 + \lambda^2)^2} \, ,
\\
&& \rho^C(r) = \frac{1}{2\pi} [1-(a+x)p(x,a)] 
= \frac{1- \sigma}{2\pi} \, .
\label{charge1}
\eea
It is readily checked with (\ref{gamma}) that both charges
are $Q^T=Q^C=+1$, however the densities do look quite
different in general. For example for the quasi--hole, $\lambda \to 0$,
we obtain from (\ref{gamma})
$\sigma = 1-\exp (-r^2/2)$, i.e.
\bea\label{hole}
\rho^T(\br) = \delta(\br) \, , 
\qquad \qquad \qquad \enspace & \lambda \to 0 \, , \\
\rho^C(\br) = \frac{1}{2\pi} \exp (-r^2/2) \, ,  
\qquad & \lambda \to 0 \, .
\label{hole1}
\eea
It is only in the opposite limit of very large skyrmions
$\lambda \to \infty$ that the two densities become similar.
But note, that although $\sigma \stackrel{\lambda \to \infty}
{\longrightarrow} 1$ in that limit, the integral
over the physical charge density, i.e. over $(1-\sigma)/2\pi$, always
remains $1$.

Thus, the conclusion of this simple exercise is twofold: (i)
$f^2 +g^2 = \sigma^2 \not= 1$, a bag is formed which does go
beyond the standard $O(3)$ description and (ii) the physical
charge density is not equal to the $O(3)$ model's topological
density, rather it is connected with the bag via (\ref{charge1}).

It should be added, that for hard-core forces at finite Zeeman 
splitting the size parameter $\lambda \to 0$ and
the (anti-) skyrmion collapses into the quasi--hole 
(\ref{hole},\ref{hole1}).
There are no skyrmions for hard-core forces. In the following we
are going to discuss the situation for Coulomb
forces.

\subsection{Coulomb interaction}

\begin{figure}[htbp]
\centerline{\epsfig{figure=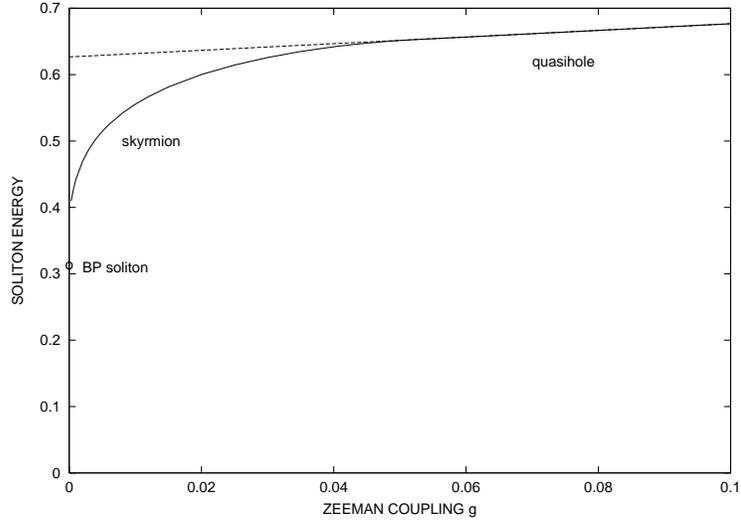,width=7cm,angle=270}}
\protect\caption{Skyrmion and quasi--hole energies
as a function of the Zeeman splitting calculated in
Hartree Fock for a pure Coulomb interaction. For $g < 0.053$ 
skyrmions are energetically favored. In the limit $g \to 0$ the
Bogomol'nyi bound is approached. 
}
\end{figure}
\begin{figure}[htbp]
\centerline{\epsfig{figure=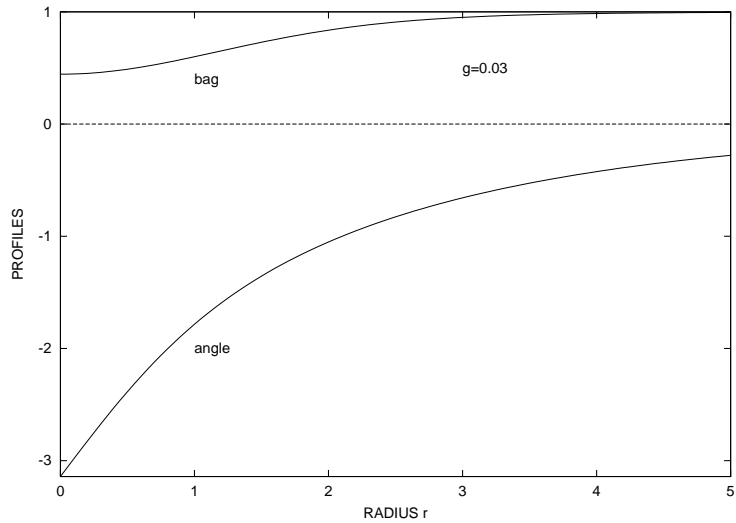,width=7cm,angle=270}}
\protect\caption{Profiles for the (anti-)skyrmion with $g=0.03$
versus the radius in magnetic lengths. A bag is formed.
}
\end{figure}
Although a realistic calculation would have to take into account 
Landau level mixing, the finite thickness of the layer 
\cite{c97,krl95,fbcdks97,mbo99}
and also
impurities, we consider the electrons in the LLL interacting
in the plane via pure Coulomb forces 
$V(\br_1-\br_2)=1/|\br_1-\br_2|$. We are interested in
bag formation and its relation to the physical charge density.
The corresponding HF equations are solved numerically using a basis
up to several thousand states for large skyrmions. The soliton energy
in dependence of the Zeeman coupling strength $g$ is 
depicted in Fig. 1. 
We observe a second order phase-transition at $g \simeq 0.053$ 
below which (anti-)skyrmions are energetically
favored compared to the quasi--hole (dashed line).
For extremely small Zeeman couplings (large skyrmions) 
the soliton energy
approaches the Bogomol'nyi bound $\sqrt{\pi/32} \simeq .3133$,
indicating that in this limit the Belavin-Polyakov soliton of
the non--linear sigma model is obtained.

The profiles are calculated microscopically via the magnetization
(\ref{fh}) and plotted in Fig. 2 for Zeeman coupling $g=0.03$. 
The formation of a bag is clearly indicated. With decreasing
Zeeman coupling the bag begins to flatten out. But
notice, that in the Belavin-Polyakov limit $g \to 0$ the
integral over $(1-\sigma)/2 \pi$ is $1$. 
\begin{figure}[t]
\centerline{\epsfig{figure=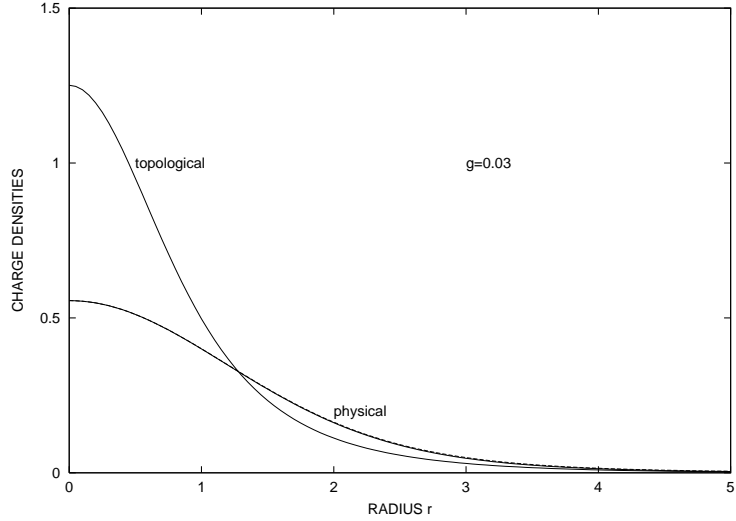,width=7cm,angle=270}}
\protect\caption{Topological and physical charge densities
for the soliton with $g=0.03$ (solid lines).
Although both total charges are normalized to $1$, the physical 
charge density
deviates significantly from the topological one. Instead it is 
extremely well approximated by the bag's shape 
$1-\sigma$ (dashed line).
}
\end{figure}
The physical charge density, calculated
microscopically according to (\ref{charge}), 
and the topological density defined via the
angle function $F$ (Fig. 2) are depicted in Fig. 3
again for Zeeman coupling $g=0.03$. It is noticed that the
physical charge density is much flatter compared to the
topological density. The bag shape function $1-\sigma$ (dashed line) 
is hardly 
distinguishable from the physical charge density. This
supports our findings for hard-core forces: it is the bag
and not the topological charge density which is closely related
to the physical charge density.

\section{Effective field theory}

There has been a lot of effort to derive an effective field theory
for the quantum Hall ferromagnet \cite{f91,mmygd95,bmv96,ab97,mm98,g99}.
The result in terms of a gradient expansion of the unit
magnetization vector $\hat{\bm}$ is essentially an $O(3)$ model,
i.e. the non--linear sigma model augmented by higher order terms
stemming from the interaction. However from the above considerations
it is expected that the full magnetization vector $\bm$,
namely its modulus $\sigma$, should show up in the effective
action. In particular the relation between the physical 
charge density and the bag (\ref{charge1}) should be respected.
The derivation of such an effective action in full
generality is a challenging task. Therefore, in the following
subsection we resort to the gradient expansion for extended
spin-textures quite as in previous approaches, however
we take bag formation into account.

\subsection{$O(3)-model$}

Starting point is the full magnetization vector and 
the physical charge density which in lowest order
gradient expansion read
\bea\label{magrad}
&& \bm = \sigma \hat{\bm} = \left( 1- 
\frac{1}{4} \partial_i \hat{\bm} \partial_i \hat{\bm} \right) 
\hat{\bm} + {\cal O} (4)
\, , \\ 
&& \rho^C = -\frac{1}{8\pi} \epsilon_{ij} 
\hat{\bm} ( \partial_i \hat{\bm} \times
\partial_j \hat{\bm} ) + {\cal O} (4)   \, .
\label{magrad1}
\eea
Because of the charge conjugation properties of the magnetization
vector, i.e. the scalar
field is the same for skyrmions and anti--skyrmions,
it is clear that in (\ref{magrad}) the energy density 
of the non--linear sigma model
rather than the topological density enters.
It is remarkable that its coefficient,
which follows e.g. from (\ref{charge1}), is independent
from the specific two-body interaction.
From the discussion in section 2 (cf. Fig. 3) it is evident that 
the physical charge density in general is not well approximated 
by the topological density, nevertheless in terms of the gradient
expansion for extended spin-textures eq. (\ref{magrad1}) holds.
In accordance with eq.(\ref{magrad}) the energy functional
could be chosen
\bea\label{effsig}
E [\bm] &=& \frac{\rho_s}{2} \int d^2r [\partial_i \bm
\partial_i \bm + (1 - \bm \bm)^2 ] + \mbox{interaction terms}
\nonumber \\
&=& \frac{\rho_s}{2} \int d^2r [\sigma^2 \partial_i \hat{\bm}
\partial_i \hat{\bm} + (1 - \sigma^2)^2 ] + \ldots \, ,  
\eea
with $\rho_s$ representing the spin--stiffness, e.g. 
$4 \pi \rho_s = \sqrt{\pi/32}$ for the Coulomb interaction.
Of course other potentials for the scalar field are possible
similar to the one derived from the QCD scaling behaviour
to describe bag formation in hadron physics \cite{gjjs86}.
However up to fourth order in the gradient expansion the
exact form of the potential will not matter provided that
it reproduces (\ref{magrad}) correctly. Note in this
context that gradients on the scalar field are of higher
order and may be neglected.

None of the previously proposed effective actions allows
for bag formation. The collective approaches
\cite{mmygd95,bmv96,ab97,g99} operate with unitary rotations
which restrict the magnetization to the unit vector $\hat{\bm}$
right from the beginning. Derivations from
Chern-Simons theories have a scalar field included which
however is not
taken seriously as a dynamical field \cite{mm98}.

In order to arrive at a non--linear $O(3)$ type of model, 
the scalar field may now be eliminated from (\ref{effsig})
in favour of a symmetric fourth order term with definite strength
related to the spin--stiffness
\be\label{effhut}
E[\hat{\bm}] = \frac{\rho_s}{2} \int d^2r 
[\partial_i \hat{\bm} \partial_i \hat{\bm}
- \frac{1}{4} ( \partial_i \hat{\bm} \partial_i \hat{\bm} )^2
+ {\cal O} (6) ] + \mbox{interaction terms} \, .
\ee
It should be emphasized that although the above $O(3)$ model
deals with a unit vector $\hat{\bm}$ the bag is 
still formed via the relation to the magnetization (\ref{magrad}).
Eqs. (\ref{magrad}) and (\ref{effhut}) are at variance with the
corresponding equations in \cite{ab97}.
The full energy functional, including interaction and Zeeman
terms is given by
\bea\label{eff}
E[\hat{\bm}] &=& \frac{\rho_s}{2} \int d^2r 
\left[ \partial_i \hat{\bm} \partial_i \hat{\bm}
- \frac{1}{4} ( \partial_i \hat{\bm} \partial_i \hat{\bm} )^2 \right]
\nonumber \\
&+& \frac{1}{2} \int \int d^2r_1 d^2r_2 \rho^C(\br_1) V(\br_1-\br_2)
\rho^C(\br_2) \\
&+& \frac{g}{4\pi} \int d^2r \left[ 1 - \left( 1- 
\frac{1}{4} \partial_i \hat{\bm} \partial_i \hat{\bm} \right) 
\hat{m_3} \right]
\, , \nonumber
\eea
with $\rho^C$ replaced by $\rho^T$ (\ref{magrad1})
in lowest order gradient expansion.
This energy functional is clearly charge conjugation invariant as
it should be. It is unique up to 
${\cal O} (4)$ in the gradients and linear
terms in the Zeeman coupling. The essential difference to 
previously proposed effective field theories is the
additional symmetric fourth order term and of course the nontrivial
connection of the $O(3)$ field with the magnetization (\ref{magrad}).

The energy functional (\ref{eff}) captures the essential features
of the microscopic approach discussed in section 2. For 
zero range forces, $4\pi \rho_s = t/4\pi$, and vanishing
Zeeman coupling it simplifies 
with the hedgehog-ansatz (\ref{magnet1}) inserted
\be\label{effhc}
E[F] = \frac{\rho_s}{2} \int d^2r 
\left[ \left( F^{\prime 2} + \frac{\sin^2 F}{r^2} \right)
- \frac{1}{4}  \left( F^{\prime 2} - \frac{\sin^2 F}{r^2} \right)^2 
\right] \, .
\ee
It is immediately verified that the Belavin-Polyakov soliton 
(\ref{fs}) solves the corresponding stability condition. With
this solution inserted the second term in (\ref{effhc})
vanishes leaving only the pure non--linear sigma model. Hence, the
soliton energy is $E=4\pi \rho_s = t/4\pi$ as in 
the microscopic calculation. Also, the charge is related
to the integral over the bag
\be
\frac{1}{2\pi} \int d^2r (1 - \sigma) 
= \frac{1}{8\pi} \int d^2r 
\left( F^{\prime 2} + \frac{\sin^2 F}{r^2} \right) = 1
\ee
as in subsection 2.1.
Again, for finite Zeeman coupling the soliton collapses. 
None of the previously proposed effective actions has this
simple property that it gives the correct result for 
extended spin--textures with hard-core forces.
It is clear that in order to obtain these results, the symmetric fourth
order term in (\ref{eff}) is essential.

In the Coulomb case,
we obtain also solitons for non--zero but weak Zeeman couplings. 
Because the repulsive interaction term scales with the inverse soliton
size $1/\lambda$ it competes with the attractive symmetric
fourth order term proportional to $-1/\lambda^2$ and there
exists a critical soliton size,
determined by the Zeeman coupling strength, below which
the soliton ceases to exist. In this way the effective
theory models qualitatively the phase-transition discussed in 
subsection 2.2. However, because of the gradient expansion employed,
it is not expected that the energy functional (\ref{eff}) remains valid
for too small solitons. Down to which soliton sizes (\ref{eff})
may give a fair description of the microscopic HF calculation
has still to be investigated.

It the 1-soliton sector it appears that the effective field 
theory has not
much advantage over the microscopic approach. The numerical
difficulty in solving the non--linear (integro-) differential 
equation for the profile function deduced from (\ref{eff}) is 
comparable to that of the HF calculation. However for 
multi--soliton systems such as skyrmion gases and crystals
\cite{cdbfgs97,nk98,mm99}
the virtues of the effective field theory become apparent.
In terms of collective coordinates the dynamics
of such multi--soliton systems may efficiently be described.
Also beyond tree approximation when loop corrections are 
considered the effective field theory approach might have advantages.
On the microscopic level this corresponds to go beyond 
HF which is complicated. 

In the following we give a brief argument why properties
like bag formation discussed in the context of the HF
calculation are not explained by loop corrections.

\subsection{Loop corrections}

Intuitively it is clear that the HF calculation on the microscopic
level should correspond to the tree approximation of the effective
theory. Accordingly, bag formation (\ref{magrad}) and the symmetric
fourth order term in (\ref{effhut}) appear already classically.
Could quantum corrections be responsible for these effects
as suggested in \cite{apfgd97}?

In the absence of an appropriate expansion parameter like
the number of colors in hadron physics the role of loop
corrections is difficult to judge. In fact, loop corrections
may lead to bag formation by softening the constraint $\hat{\bm}^2 =1$.
Also the energy is lowered and the charge distribution broadened.
Using heat kernel expansion methods one can even show that
loop corrections effectively lead to a symmetric fourth order
term with the correct (attractive) sign. Nevertheless, we 
argue that these effects observed in the HF calculation
cannot be attributed to loop corrections for the subsequent
argument.

Let us for simplicity consider only the non--linear 
sigma model part in the
effective action. The equations for the fluctuations are independent
of the strength of that term, i.e. the spin--stiffness $\rho_s$.
The only dimensionful parameter which appears is the
size $\lambda$ of the BP soliton. For that reason the effective
symmetric fourth order term produced by the heat kernel expansion
comes with a coefficient proportional to $\lambda$ such that
the total term scales like $1/\lambda$. In fact, the whole Casimir energy
scales like $1/\lambda$ as was explicitely demonstrated in \cite{wh00}. 
In any case, the spin--stiffness does not appear at all, which implies
that the symmetric fourth order term found in (\ref{effhut})
necessary to model the microscopic HF calculation 
cannot be explained by quantum fluctuations. From that
argument it becomes clear that the bag is already formed in tree level
according to (\ref{magrad}) and (\ref{effhut}).

The above reasoning  does not imply that fluctuations are unimportant.
In fact, we expect significant loop corrections \cite{wh00},
which among other things also will intensify the bag formation. 
But in the microscopic picture this would correspond
to go beyond HF \cite{krl95} performing e.g. a RPA calculation.

\section{Conclusions}

Charged skyrmions in quantum Hall ferromagnets at filling
factor $\nu=1$ were reanalysed in the framework of the HF
theory. It was demonstrated, that within the LLL approximation
the hedgehog
ansatz leads to the lowest energy configuration in the unit
charge sector. Calculating all three components of the 
magnetization vector it was noticed that its length deviates
locally from unity, i.e. a bag is formed which accommodates
the excess charge. The physical charge density is extremely
well approximated by the bag, but in general deviates 
considerably from the topological density. Hence, the
commonly accepted and celebrated equivalence of physical and 
topological charge does not hold locally.

In terms of a gradient expansion an improved $O(3)$ type of 
effective action is proposed for the
quantum Hall ferromagnet, which takes bag formation
into account. Accordingly, the magnetization vector
is nontrivially related to the $O(3)$ unit vector field.
In addition to previous approaches, the effective action
contains an attractive symmetric fourth order term.
For hard-core forces it was explicitely demonstrated
that with these amendments made the correct microscopic
result for extended spin--textures is obtained.
For more realistic forces the effective action models
the phase--transition anti--skyrmion to quasi--hole
(skyrmion to quasi--particle) observed in the
microscopic calculation when the Zeeman splitting
is increased. 
The additional fourth order term also modifies the 
soliton--soliton interaction, e.g. it influences
the dipole--strength in the dipole--dipole force
for skyrmions. By this means  the properties of 
multi--skyrmion systems such as skyrmion gas and
lattice are affected.

The present investigation can be completed in several
respects. The range of validity of the proposed
energy functional for realistic two--body forces
may carefully be tested by comparison with the
microscopic HF calculation. Also the time--derivative
part of the effective action \cite{ab97} is of
interest and will be subject to changes due to the
bag. Beyond that, the derivation of a corresponding
effective field theory without resorting to the
gradient expansion is desirable. Such a theory should,
in addition to the $O(3)$ unit vector field, contain
a scalar field for the magnetization vector's
length as a dynamical field. Finally, loop
corrections may be calculated along the lines of
ref. \cite{wh00}. On the microscopic level this
can be complemented by RPA type calculations
which take fluctuations into account.

\section{Acknowledgments}

The author wishes to thank his colleagues G. Holzwarth,
T. Flie\ss bach, F. Meier and A.-S. Marculescu for 
contributing to a seminar about
the quantum Hall effect from which this study has emerged. 
This work was supported in part by funds provided by the
Deutsche Forschungsgemeinschaft (Contract No. DFG-Ho-527/16-1).

\end{document}